\documentclass[conference]{IEEEtran}
\IEEEoverridecommandlockouts

\usepackage{cite}
\usepackage{amsmath,amssymb,amsfonts}
\usepackage{algorithmic}
\usepackage{graphicx}
\usepackage{textcomp}
\usepackage{xcolor,threeparttable,booktabs,multirow,url,subcaption}
\usepackage{authblk}
\def\BibTeX{{\rm B\kern-.05em{\sc i\kern-.025em b}\kern-.08em
    T\kern-.1667em\lower.7ex\hbox{E}\kern-.125emX}}

\begin{document}

\title{HAP: Hybrid Adaptive Parallelism for Efficient Mixture-of-Experts Inference
}

\author[1]{Haoran Lin}
\author[2]{Xianzhi Yu}
\author[2]{Kang Zhao}
\author[2]{Han Bao}
\author[2]{Zongyuan Zhan}
\author[2]{Ting Hu$^2$\\
{Wulong Liu}}
\author[1]{Zekun Yin}
\author[*]{Xin Li$^1$}
\author[*]{Weiguo Liu$^1$}
\affil[1]{School of Software, Shandong University, Jinan, China}
\affil[2]{Huawei Noah’s Ark Lab, Beijing, China}
\affil[ ]{Email: haoran.lin@mail.sdu.edu.cn}

\maketitle

\begin{abstract}
Current inference systems for Mixture-of-Experts (MoE) models primarily employ static parallelization strategies. However, these static approaches cannot consistently achieve optimal performance across different inference scenarios, as they lack the flexibility to adapt to varying computational requirements. In this work, we propose HAP (Hybrid Adaptive Parallelism), a novel method that dynamically selects hybrid parallel strategies to enhance MoE inference efficiency. The fundamental innovation of HAP lies in hierarchically decomposing MoE architectures into two distinct computational modules: the Attention module and the Expert module, each augmented with a specialized inference latency simulation model. This decomposition promotes the construction of a comprehensive search space for seeking model parallel strategies. By leveraging Integer Linear Programming (ILP), HAP could solve the optimal hybrid parallel configurations to maximize inference efficiency under varying computational constraints. 
Our experiments demonstrate that HAP consistently determines parallel configurations that achieve comparable or superior performance to the TP strategy prevalent in mainstream inference systems. 
Compared to the TP-based inference, HAP-based inference achieves speedups of 1.68$\times$, 1.77$\times$, and 1.57$\times$ on A100, A6000, and V100 GPU platforms, respectively. 
Furthermore, HAP showcases remarkable generalization capability, maintaining performance effectiveness across diverse MoE model configurations, including Mixtral and Qwen series models.
\end{abstract}


\section{Introduction} 
Recently, Mixture-of-Experts (MoE) large language models (LLMs) have gained popularity in achieving state-of-the-art performance while maintaining computational efficiency, making them increasingly pivotal in real-world applications, such as language modeling, machine translation, and image recognition \cite{artetxe2021efficient,costa2022no,fedus2022switch,liu2024deepseek,riquelme2021scaling}.
To support their deployment, modern inference systems like vLLM and DeepSpeed-FastGen have emerged, showcasing remarkable capabilities in managing the complexity of MoE models \cite{kwon2023efficient,holmes2024deepspeed,10528887} . However, these systems predominantly rely on static parallelization strategies, such as Tensor Parallelism (TP) and Expert Parallelism (EP) \cite{lepikhin2020gshard}, which suffer from critical limitations in adapting to dynamic inference scenarios.  
 
We identify three fundamental challenges associated with static parallel strategies, in aspects of computation, communication, and model architecture.
1) \textbf{Unflexible tensor partition}. 
Static strategies may result in suboptimal performance in some inference scenarios, as the fixed tensor partition fails to fully leverage the computational capabilities of the hardware for specific operators. Conversely, dynamically switching parallel strategies during inference incurs significant communication overhead, potentially offsetting any prospective performance improvements.
2) \textbf{Mismatched bandwidth utilization of communication}. Parallel strategies inherently impose distinct communication patterns, such as All-to-All for EP while AllReduce for TP. 
Static optimization strategies cannot dynamically adapt to these distinct pattern requirements, potentially exacerbating communication overhead, especially in heterogeneous hardware environments.  
3) \textbf{Inefficiency in addressing diverse MoE architectures}. 
MoE models with different expert configurations, such as varying expert counts and distinct shared expert implementations, require scenario-specific computation/communication patterns. Static strategies may struggle to adapt to model-specific configurations, leading to suboptimal performance.

To address these limitations, we propose Hybrid Adaptive Parallelism (HAP), a novel method that automatically selects hybrid parallel strategies tailored to both model architectures and hardware capabilities. 
HAP introduces two key innovations: \textbf{Module Decomposition} and \textbf{Hybrid Adaptive Parallel Strategies}.
Module Decomposition involves decomposing MoE models into Attention and Expert modules, each equipped with a dedicated inference latency simulation model to enable fine-grained inference simulation. Hybrid Adaptive Parallel Strategies constructs a hierarchical search space and leverages Integer Linear Programming (ILP) to dynamically identify the optimal combination of parallel strategies for Attention and Expert modules while adhering to hardware constraints.

Our contributions are summarized as follows:  
\begin{itemize}
    \item We develop module-specific simulation models that precisely capture computational and communication costs, enabling fine-grained performance prediction. \item We design a novel dynamic parallelism transition strategy to minimize the overhead associated with switching parallel configurations during inference.
    \item HAP formulates parallel strategy selection as an ILP problem, efficiently exploring a combinatorial search space to maximize device utilization and minimize end-to-end latency.                                                                  
    \item Evaluations across diverse MoE architectures and GPU platforms demonstrate the generalization capability of our HAP, which achieves up to 1.77$\times$ speedup when integrated with DeepSpeed-FastGen.  
\end{itemize}

\section{Related Work}
\subsection{MoE Model Architectures}
\begin{figure}[htbp]
\includegraphics[width=1\linewidth]{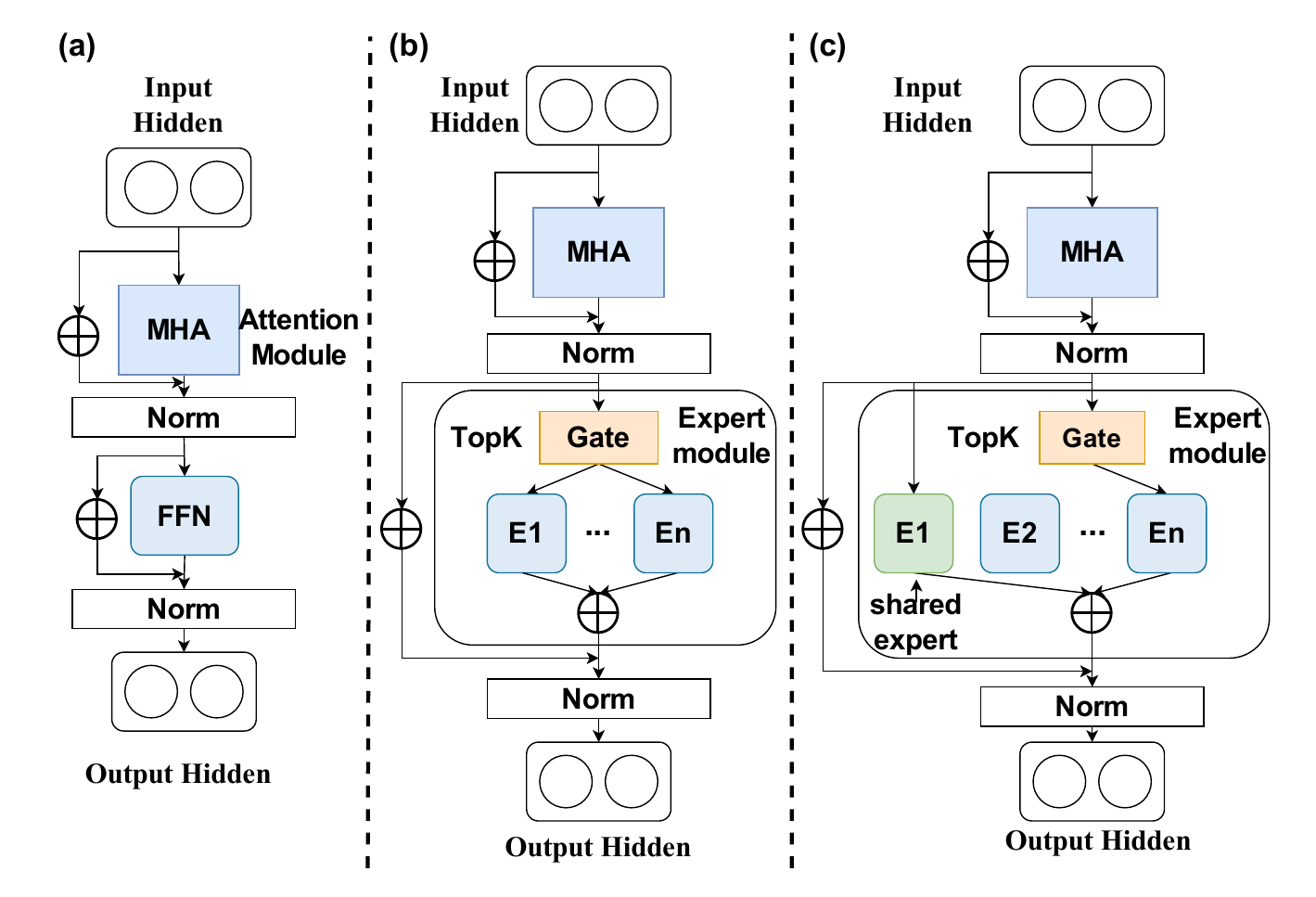}
\caption{(a) The typical dense transformer layer consists of an Attention module followed by a Feed-Forward Network (FFN) module.
(b) The naive MoE transformer layer replaces the FFN module with an Expert module. 
 (c) The MoE Transformer layer with shared experts incorporates experts that are always activated, known as shared experts. 
 } 
\label{fig:moe-arc}
\end{figure}

As illustrated in Fig.~\ref{fig:moe-arc} (b), MoE models replace the standard feed-forward network (FFN) module with an Expert module comprising multiple parallel expert networks and a gating network, which employs a Top-K selection mechanism to dynamically dispatch input tokens to the most relevant experts \cite{zoph2022designing,jiang2024mixtral}. 
Notably, some advanced MoE models introduce shared experts \cite{qwen_moe,liu2024deepseek,yang2024qwen2technicalreport}, which simultaneously reduce parameter redundancy and enhance cross-layer knowledge transfer, as shown in Fig.~\ref{fig:moe-arc}(c). 
During inference, MoE models activate only a small subset of experts per token, thereby decoupling the growth of FLOPs from the parameter count. 

\vspace{-10pt}
\subsection{Generative Inference}
Generative inference consists of two distinct computational stages: the prefill stage and the decoding stage, each exhibiting unique performance characteristics. The prefill stage processes the entire input prompt, exhibiting compute-bound characteristics due to its intensive tensor computations. 
Conversely, the decoding stage generates tokens sequentially, with each token generation dependent on previously generated tokens. This stage typically involves frequent loading of model weights and KV cache, leading to memory-bound behavior. 

\subsection{Static Parallelism} 
\paragraph{\textbf{Data Parallelism}} (DP) replicates the entire model across devices while partitioning input data batches. In batch-oriented inference scenarios, DP eliminates communication overhead, enabling near-linear throughput scaling across distributed systems. 
However, the replication of models results in redundant memory usage, which consequently necessitates the integration of complementary parallelism strategies to alleviate the memory constraints.

\paragraph{\textbf{Tensor Parallelism}} (TP) partitions model weights across multiple devices, enabling the deployment of large language models (LLMs) that exceed the memory capacity of a single device. It is widely supported by most mainstream inference frameworks, such as TensorRT-LLM, vLLM, Deepspeed-FastGen, and SGLang \cite{nvidia2024tensorrt,kwon2023efficient,holmes2024deepspeed,zheng2024sglang}, making it the most commonly adopted static parallelism approach in LLM inference. However, TP incurs significant communication overhead, which can degrade throughput and scalability \cite{li2024flash}.

\paragraph{\textbf{Expert Parallelism}} (EP) distributes experts across multiple devices, achieving superior scalability and computational efficiency \cite{huang2023towards}.
However, the frequent communication may become a performance bottleneck during inference. Furthermore, EP may lead to load imbalance, where certain devices are overutilized while others remain underutilized, reducing overall resource efficiency. Nowadays, DeepSpeed-MoE employs a static hybrid parallel strategy integrating DP, TP, and EP, demonstrating superior performance in large-scale MoE model inference \cite{rajbhandari2022deepspeed,singh2023hybrid}. DeepEP is tailored for MoE inference with static EP, offering high-throughput and low-latency all-to-all GPU kernels, thereby achieving efficient inference \cite{deepep2025}.

\paragraph{\textbf{Pipeline Parallelism}} (PP) distributes layers across multiple devices for parallel computation, thereby achieving superior scalability. However, this approach introduces pipeline bubbles that can significantly degrade inference throughput. Given the inherent performance limitation, we deliberately exclude pipeline parallelism from the architectural search space in our work.

\vspace{-5pt}
\subsection{The adaptive parallelism}
Alpa \cite{zheng2022alpa} leverages ILP to systematically identify optimal intra-node parallelism configurations, thereby providing an effective adaptive methodology. 
However, this approach exhibits significant limitations by exclusively focusing on communication latency while disregarding computational latency during its search process. Consequently, the derived parallelization strategies are not applicable for inference tasks, which is not selected as comparisons for our work.
Moreover, Tutel \cite{hwang2023tutel} introduces a switchable parallelism mechanism specifically designed for training scenarios, which meticulously analyzes the communication complexity of various parallel strategies within the Expert module and implements the most optimal strategy. Nevertheless, Tutel cannot achieve zero-cost switchable parallelism during inference and fail to implement appropriate parallel strategies for both the Attention and Expert modules simultaneously.

\section{Methodology}
\subsection{Analysis of MoE Inference}
\label{sec:analysis}
\subsubsection{Runtime Breakdown of MoE Inference.}
Execution time (i.e., latency) is a critical metric in model inference, as shorter latency ensures more timely service responses. 
In this subsection, we present a detailed performance breakdown of MoE inference with respect to latency. The evaluation is conducted on a node equipped with 4 A6000 GPUs, with a sequence length set to 2K. We compare the execution times of both the prefill stage and the decoding stage under various parallel strategies.

\begin{figure}
    \centering
    \includegraphics[width=1\linewidth]{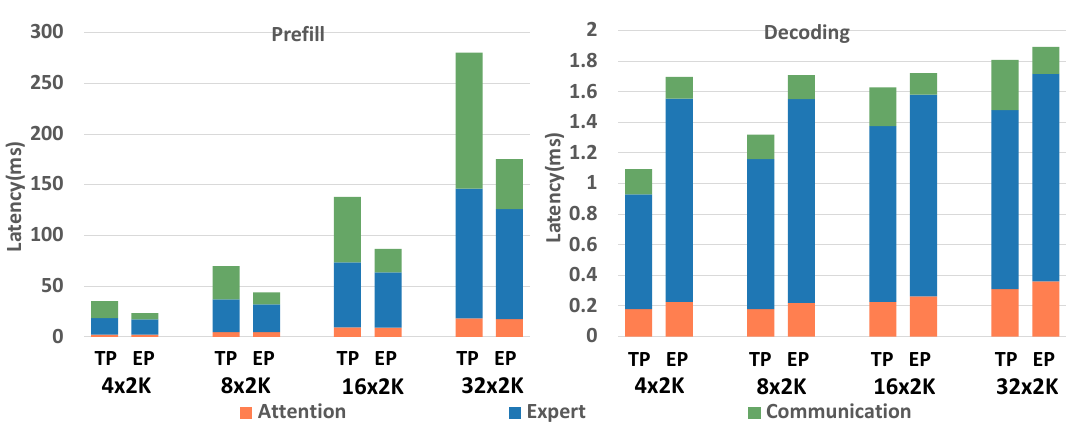}
    \caption{The per-layer latency breakdown during both prefill and decoding stages of Mixtral-8x7B inference, evaluated under TP and EP configurations across 4 NVIDIA A6000 GPUs.}
    \label{fig:analysis}
\end{figure}

We decompose the inference process into three principal components: the computation of the Attention module, the computation of the Expert module, and collective communication. As illustrated in Fig.~\ref{fig:analysis}, the latency differences between the TP and EP strategies are primarily driven by the computational cost of the Expert module and the overhead associated with collective communication.  Specifically, intra-node communication is executed via PCIe \cite{9231458}, which is characterized by limited bandwidth, resulting in high sensitivity to data volume. Consequently, during the prefill stage, TP incurs higher communication latency compared to EP due to its larger communication volume, resulting in degraded inference performance. 
In the decoding stage, where the communication volume is generally minimal, the load imbalance introduced by EP leads to inefficient computation of the Expert module compared to TP. 

These observations suggest that when selecting parallel strategies for the Expert module, scenarios with low intra-node communication bandwidth, where communication constitutes a performance bottleneck, should favor strategies with lower communication volume, whereas environments with high intra-node communication bandwidth (e.g., A100 GPU nodes utilizing NVLink) \cite{choquette2020nvidia} should adopt strategies that prioritize computational efficiency. Moreover, employing distinct parallel strategies for the prefill and decoding stages typically necessitates the redistribution of model weights through collective communication, \textbf{which introduces substantial overhead and underscores the demand for more efficient solutions.}

\subsubsection{Memory Consumption.}
The memory requirements are primarily attributed to three components: model weights, the KV cache, and activations. 
We analyze the memory consumption of the Attention module and the Expert module separately. 
For the Attention module, when employing TP, DP, or a combination of DP and TP, the per-device memory consumption of the KV cache and activations remains consistent across these strategies. 
However, \textbf{DP demonstrates a $d \times$ increase in memory consumption of model weight relative to TP}, where $d$ denotes the DP degree.
In contrast, Expert modules exhibit identical per-device model weight memory footprints regardless of parallel strategies. 
Current state-of-the-art (SOTA) work such as DeepEP employ non-uniform All-to-All communication operations to manage expert-parallel data dispatch and combination. \textbf{This approach introduces different activation memory consumption across devices due to imbalanced expert workload distribution.} 
Consequently, the HAP should take the memory constraints incurred by DP and EP into accounts.
To conservatively estimate this memory requirement for EP, we adopt an empirical upper-bound activation memory approximation that doubles the baseline TP activation footprint. This bounding methodology accounts for potential workload imbalances while maintaining analytical tractability.

We design the method named HAP to automatically search the optimal hybrid parallel strategies: 
1) We develop inference simulation models for the Attention module, Expert module, and collective communication operations, enabling the accurate estimation of the inference latency.
2) By leveraging these simulation models to quantitatively evaluate the end-to-end inference latency for various strategy combinations, we build a search space that encompasses all feasible parallelization strategies for both modules.
3) Aimed at minimizing total inference latency, we formulate and simplify the problem as Integer Linear Programming (ILP), allowing for the efficient identification of the optimal hybrid parallelization strategy. 

\subsection{Inference Latency Estimation}
To ensure the effectiveness of optimal strategy search, the simulation of inference latency must exhibit accuracy and stability. This subsection presents our inference latency simulation model. 

We split the inference duration into several items to make an accurate simulation. Specifically, the total inference duration is represented as equation \ref{eq1}.
\begin{align}
\label{eq1}
T_{total} &= T_{prefill} + T_{decoding} \\
T_{prefill} &= N_{layer} * (T_{attn} + T_{experts} + T_{comm})\\
T_{decoding} &= S_{output} * N_{layer} * (T_{attn} + \nonumber \\
             &\quad  T_{experts} + T_{comm})
\end{align}
The temporal parameters $T_{attn}$, $T_{experts}$, and $T_{comm}$ are obtained from inference simulation models. $T_{total}$ denotes the overall inference time, with $T_{prefill}$ and $T_{decoding}$ representing the latencies of the prefill and decoding stages. $N_{layer}$ is the number of model layers, and $S_{output}$ is the output sequence length. Within a single layer, $T_{attn}$, $T_{experts}$, and $T_{comm}$ correspond to the latencies of the attention module, expert module, and communication, respectively.

The simulation models for the Attention and Expert modules are computational estimation model based on Floating-Point Operations (FLOPs). We formulate the simulation model as: $T_{cal} = \left( \frac{F_{module}}{Max\_FLOPs/s} \right) \times \eta$.
$F_{module}$ denotes the number of FLOPs for a given module and ${Max\_FLOPs/s}$ represents the device’s theoretical peak FLOPS.
We employ the model configuration and input prompt information, i.e., b,s,h, to parameterize $\eta$. Additionally, these parameters are enriched through polynomial feature expansion to enhance representational capacity, which is then leveraged by an efficient random forest regression model to fit $\eta$. The lightweight architecture of the regression model ensures minimal computational overhead while achieving high predictive accuracy.

Similar to the computational simulation model, the communication simulation model is carried out by the following principle: $ T_{comm} = \left( \frac{V_{data}}{Bandwidth} \right) \times \rho$.
$V_{data}$ is the data volume in the collective communication and $Bandwidth$ implys the network bandwidth within the server.
We also utilize a random forest regression model to estimate $\rho$, which exclusively takes the data volume and the device bandwidth as its input. Notably, the training datasets derive from empirically measured operator runtime latency values, acquired through systematic benchmarking protocols. 

\subsection{Optimal Hybrid Parallel Strategies Search}
We summarize all possible parallelization strategies for both the Attention and Expert modules:
\begin{itemize}
    \item \textbf{Attention module.} The potential deployment strategies for the Attention module weights include DP, TP, and the hybrid parallel strategies combining both DP and TP.
    \item \textbf{Expert module.} The deployment strategies for the Expert module weights encompass EP, TP, and the hybrid parallel strategies that combine both EP and TP. Given that the Expert module weights constitute the majority of the parameters in MoE models and considering memory constraints, we exclude DP for this module from the search space.
\end{itemize}
Notably, for the aforementioned hybrid parallel strategies, the TP degree increases exponentially as a power of two. The combination of these strategies constitutes the search space. The objective of searching the optimal parallel strategy is to minimize inference latency. We formulate this cost minimization problem as an ILP problem. The formal definition of the integer programming problem is as follows:

\begin{equation}
\begin{split}
    \min_{S,E,R} \ & N_{\text{layer}} \times \{ 
    \underbrace{S_k^T T_{a} + E_i T_{e} +T_{C_{ki}}}_{\text{prefill}} 
    + S_{\text{output}} \times \\
    & \underbrace{(S_k^T T_{a} + E_j T_{e} +T_{C_{kj}})}_{\text{decode}} \}+ \underbrace{E_i^T \mathbf{C_{ij}} E_j}_{\text{switching cost}}
\end{split}
\end{equation}

\begin{equation}
\begin{aligned}
\text{s.t.} \quad & A_t, A_d, E_d, E_t, E_e \in \mathbb{N}^+ \\
& \frac{M_{KV} + A_d \times M_{attn} + E_d \times M_{exp}}{N} + 2M_{act} < M_{gpu}\\
& N = A_t \times A_d = E_d \times E_t \times E_e \\
& B = b \times A_d \\
& Dim \mid A_t, N_{kv} \mid A_t, N_{experts} \mid E_e, Dim_{exp} \mid E_t
\end{aligned}
\end{equation}

Specifically, the $a \mid b$ represents a divides b. $N_{\text{layer}}$ is the number of model layers, and $S_{\text{output}}$ is the output length. 
Let $A_t$ and $A_d$ denote the TP and DP degrees for the Attention module, while $E_t$, $E_d$ and $E_e$ represent the TP, DP and EP degrees for the Expert module, respectively.
$M_{KV}$, $M_{attn}$, $M_{exp}$, and $M_{act}$ indicate the memory usage of the KV cache, attention weights, expert weights, and activations, respectively, while $M_{gpu}$ is GPU memory capacity. $B$ denotes the batch size, $b$ is the micro-batch size per DP device, and $N$ is the total number of devices.$Dim$ is the hidden size, $Dim_{exp}$ is the expert's intermediate size, and $N_{experts}$ denotes the number of experts per layer. The number of parallel strategies for the Attention module and the Expert module are are denoted by $K_a$ and $K_e$,respectively. The computational cost of the Attention module is represented by the vector $T_a$ of length $K_a$, where $T_{ai}$ denotes the computational cost of the $i$-th parallel strategy for the Attention module. Similarly, the Expert module has a computational cost vector $T_e$. The communication cost associated with the selected parallel strategies for both modules is denoted as $T_C$. 

Moreover, we define the one-hot decision vectors $ S_k \in \{0,1\}^{K_a}$ and $E_i \in \{0,1\}^{K_{e}}$ to indicate the chosen parallelization strategies, where $S_{ki} = 1$ indicates that the $i$-th parallel strategy is selected for the Attention module. Additionally, the matrix $\mathbf{C} \in \mathbf{R}^{K_e \times K_e}$ represents the overhead incurred when switching the Expert module's parallelization strategy from the prefill stage to the decoding stage, with $\mathbf{C_{ij}}$ denoting the overhead of switching from the $i$-th strategy to the $j$-th strategy.
Furthermore, we prune the search space based on memory requirements and prior empirical knowledge. 
Due to the KV cache, the Attention module employs a consistent parallelization strategy across both prefill and decoding stages, while the Expert module can adopt different strategies for each stage.
Based on memory constraints, we preemptively eliminate some parallel strategies for the Expert module that utilize DP. Leveraging existing experience, we exclude combinations of DP, EP, and TP in the parallel strategies of Expert module for the generally suboptimal performance on a node with multiple devices. 

We formulate and solve the ILP problem using Python's PuLP library. For typical limited-scale deployment scenarios (e.g., single-machine 8-GPU configurations), the optimization completes consistently within one second of search time. Crucially, our experimental methodology incorporates the ILP solver runtime into all reported end-to-end latency measurements across benchmark evaluations.

\begin{table*}[htbp]
\centering
\small
\caption{Performance comparison of different quantization schemes with Mixtral 8x7B.}
\begin{tabular}{lcccccccccccc}
\toprule
\textbf{Quantization scheme}  & \textbf{MMLU}  & \textbf{Arc-e} & \textbf{Arc-c} & \textbf{NQ\_OPEN} & \textbf{TriQA} & \textbf{HumanE} & \textbf{MBPP}  & \textbf{GSM8K} \\
\midrule
\textbf{Origin} &67.7\% &   84.3\% & 56.9\% & 30.2\% & 71.4\% & 35.9\% & 47.6\%  & 58.3\%  \\
\midrule
\textbf{Per-tensor}      & 67.7\%  & 83.3\%  & 56.4\% & 29.6\% & 70.2\% & 35.9\% & 47.8\% & 55.6\% \\
\textbf{Per-group}    & 67.7\% &   84.0\% & 56.2\% & 30.1\% & 70.6\% & 35.4\% & 47.6\%  & 58.0\% \\
\bottomrule
\end{tabular}
\label{tab:quant_vs_origin}
\end{table*}

\subsection{The Dynamic Parallelism Transition Strategy.} 
\begin{figure}[htbp]
    \centering
    \includegraphics[width=0.9\linewidth]{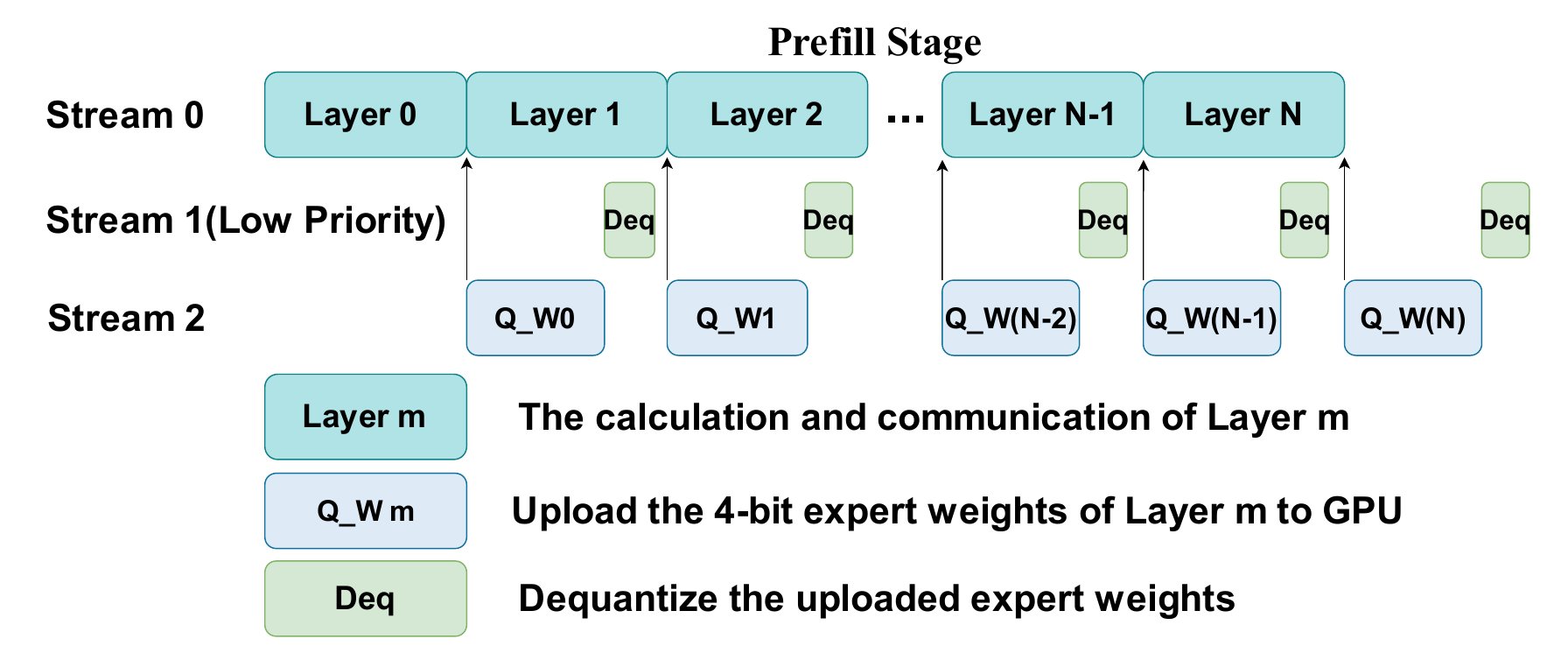}
    \caption{The workflow of parallelism transition during the prefill stage.}
    \label{fig:quant_deq}
\end{figure}

Parallelism strategy transitions typically necessitate redistributing model weights via collective communication operations, such as AllGather and AllToAll, which can incur substantial communication overhead. This challenge arises due to the expert module weights, which account for approximately 90\% of the total model parameters. As a result, naïve parallelism transitions may paradoxically degrade inference performance. To address this limitation, we propose a dynamic parallelism transition method with two implementation approaches: (1) Weight redistribution via collective communication; 
(2) As shown in Fig.~\ref{fig:quant_deq}, we maintain a 4-bit (INT4) quantized backup of expert layer weights in CPU memory. This backup is asynchronously uploaded to the corresponding devices through multi-stream pipelines and subsequently dequantized to its native precision, e.g., BF16 \cite{bitsandbytes2024}. The optimal transition approach is selected through empirical simulations:
\begin{equation}
    \begin{split}
    \mathbf{C_{ij}} &= \min \{ T_{reshard}, \max \{0, T_{upload} \\
    & + T_{dequant} - (S_k^T T_{a} + E_i T_{e} +T_{C_{ki}})\}\}
    \end{split}
\end{equation}
where $T_{reshard}$ denotes the time of weight redistribution via collective communication operations, and $T_{upload}$ represents the time taken to upload weights from the CPU to the devices. The overhead of weight dequantization is denoted by $T_{dequant}$. Note that $T_{dequant}$ scales with the number of dequantized parameters ($V_{dequant}$). Accordingly, we determine $V_{dequant}$ based on the number of GPUs, thereby constructing a dictionary of $V_{dequant} \rightarrow T_{dequant}$, which is queried at runtime to obtain this overhead.

\vspace{-10pt}
\vspace{-10pt}

Notably, expert weight quantization followed by dequantization maintains $>$99.5\% cosine similarity to original weights, yet can still cause performance degradation. As demonstrated in Table~\ref{tab:quant_vs_origin}, per-tensor quantization significantly degraded GSM8K performance while remaining nearly lossless on MMLU and MBPP. Through comprehensive evaluation of per-tensor, per-channel, and per-group quantization \cite{nagel2021white}, we dopted fine-grained per-group quantization, which preserved model performance with only marginal degradation.


\section{Performance Evaluation}
\subsection{Overview}
As shown in Table~\ref{tab:exp_config}, we conducted systematic experiments across four orthogonal inference scenarios defined along two critical dimensions: input context scale (short versus long) and output generation complexity (constrained versus extended). Utilizing the open-source framework DeepSpeed-FastGen as our experimental platform, we performed comparative evaluations of end-to-end latency between our proposed HAP optimized inference and TP-based inference, with the latter serving as the baseline configuration.

\begin{table}[htbp]
\caption{The experimental configurations used in the evaluation.}
\centering
\resizebox{1\linewidth}{!}{
\begin{tabular}{l|l}
\hline
Inference scenarios                         & Configurations                            \\ \hline
Short-context input with constrained output & 256-token context, 64-token generation    \\
Short-context input with extended output    & 256-token context, 2048-token generation  \\
Long-context input with constrained output  & 4096-token context, 64-token generation   \\
Long-context input with extended output     & 4096-token context, 2048-token generation \\ \hline
\end{tabular}
}
\label{tab:exp_config}
\end{table}

We conduct extensive evaluations using HAP on A6000, A100 and V100 GPUs. The A6000 and V100 GPUs execute the intra-node communication via PCIe while A100 via NVLink. We employ the Mixtral-series and Qwen-series MoE models to demonstrate the superior performance and generalizability of HAP. The Mixtral-series features a smaller number of experts per layer, each with a larger parameter count. In contrast, the Qwen-series incorporates a larger number of experts per layer, each with fewer parameters, and additionally supports shared experts. Table~\ref{tab::model_config} elaborates the model configurations. 

\begin{table}[htbp]
\caption{The model configurations used for the inference performance evaluation.}
\centering
\setlength{\tabcolsep}{8pt} 
\begin{threeparttable}
\let\cline\cmidrule
\resizebox{1\linewidth}{!}{
\begin{tabular}{ccccccc}
\toprule
\multirow{1}{*}{Model}  & \multicolumn{1}{c}{Params(B)}  & \multicolumn{1}{c}{Layers} & \multicolumn{1}{c}{Q\_Heads} & \multicolumn{1}{c}{Hidden} & \multicolumn{1}{c}{Experts} & \multicolumn{1}{c}{MoE\_inter\_size}\\
\midrule
Mixtral-8x7B  & 46.7 & 32 & 32  & 4096 & 8 & 14336\\
Qwen1.5-MoE-A2.7B    & 14.3 & 24 & 16  & 2048 & 60 & 1408\\
Qwen2-57B-A14B  & 57.4  & 28 & 28 & 3584 & 64 &2560\\
\bottomrule
\end{tabular}
}
\end{threeparttable}
\label{tab::model_config}
\end{table}

\begin{figure*}[htbp]
    \centering
    \includegraphics[width=0.9\linewidth]{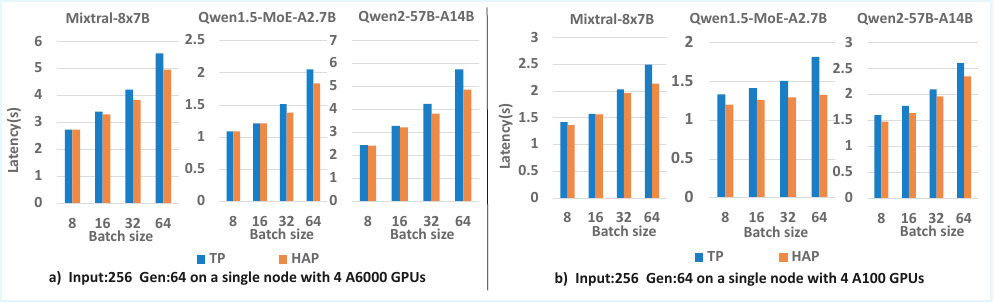}
    \caption{The performance comparison conducted under a scenario featuring a 256-token context and 64-token generation between HAP and the TP baseline, with evaluations performed on Mixtral-series and Qwen-series MoE models using 4×A6000 GPUs and 4×A100 GPUs, respectively.}
    \label{fig:256-64}
\end{figure*}
\vspace{-10pt}
\subsection{The Estimation of Simulation Models.}
\begin{figure}[!htbp]
  \centering 
  \begin{subfigure}[t]{0.8\linewidth} 
    \includegraphics[width=\linewidth]{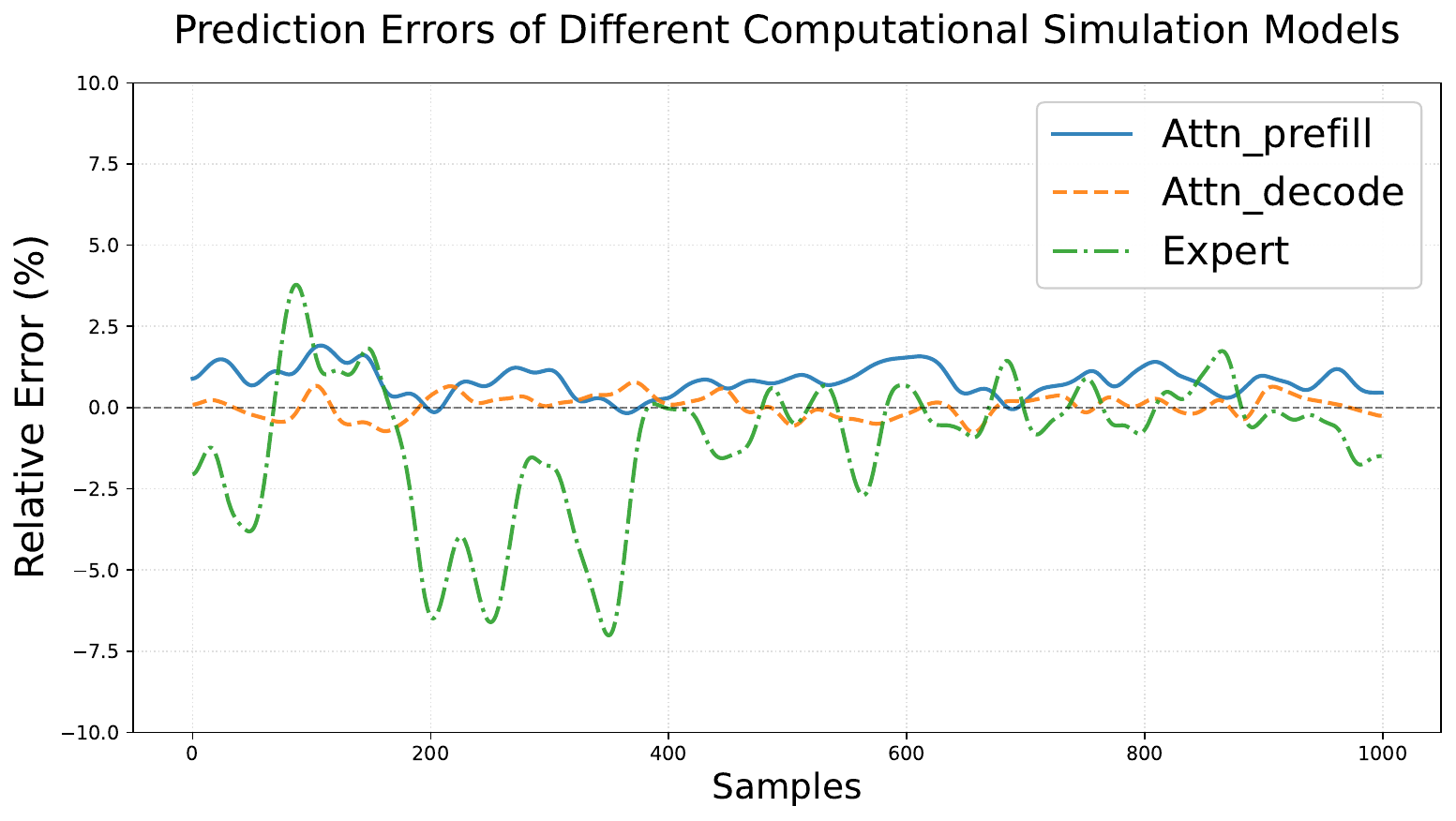} 
    \label{fig:calcul} 
  \end{subfigure}
\vspace{-20pt}
  \begin{subfigure}[t]{0.8\linewidth}
    \includegraphics[width=\linewidth]{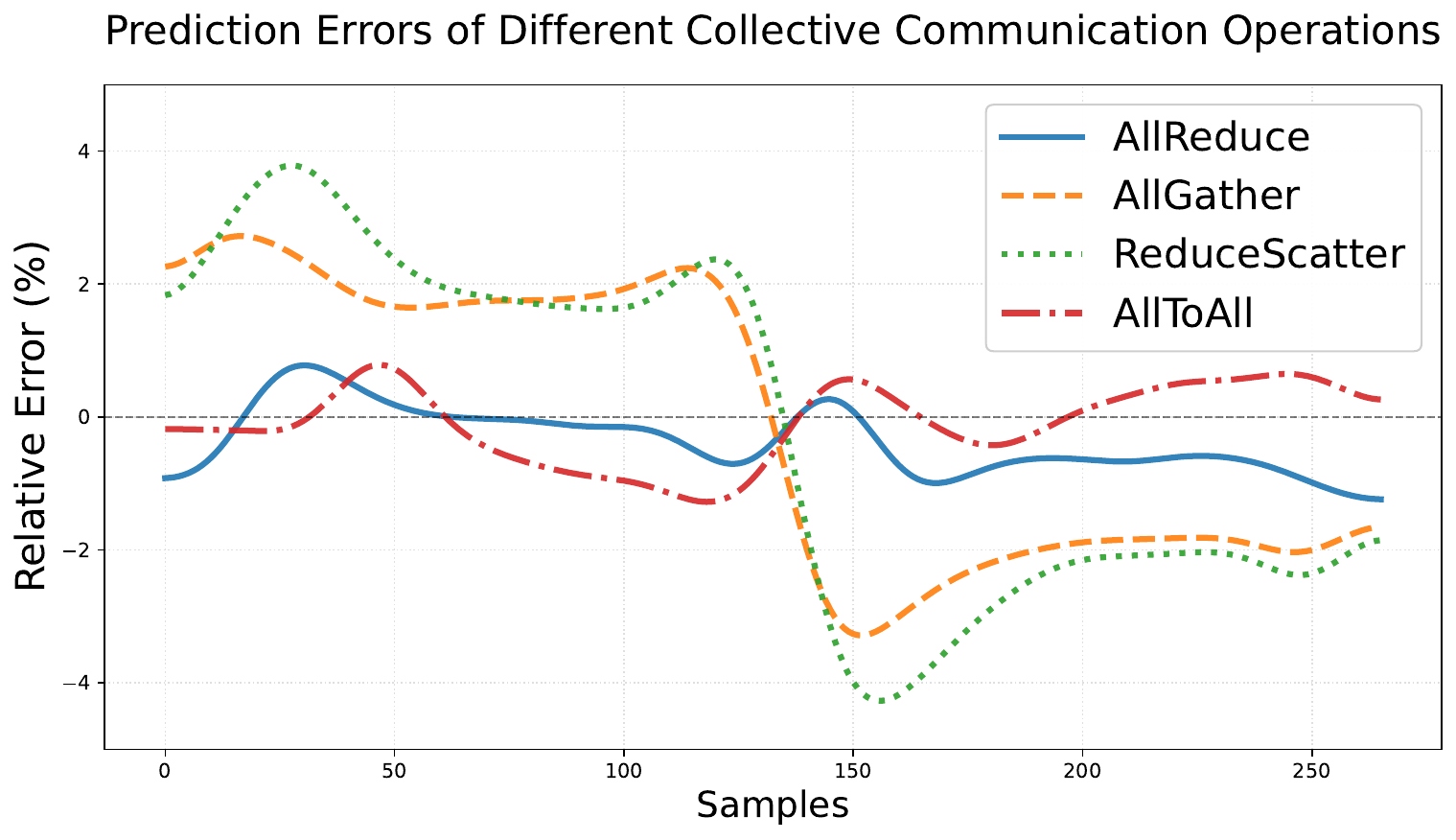}
    \label{fig:communi}
  \end{subfigure}
  \caption{Comparative evaluation of numerical accuracy in computational and communication simulation models}
  \label{fig:full} 
\end{figure}

We constructed the test datasets using empirically measured computation and communication latency profiles obtained during the inference process. These datasets were subsequently utilized to assess the prediction accuracy of simulation models. As illustrated in Fig.~\ref{fig:full}, the communication simulation models exhibit an error margin within 5\%, while the computational simulation models maintain an error below 10\%. This level of accuracy is sufficient to support the construction of effective and reliable search spaces.

\subsection{ Performance details of HAP}
\subsubsection{Short-context input with constrained generation.}
Compared to the baseline, Fig.~\ref{fig:256-64} shows that the HAP-based inference achieved maximum speedups of 1.13$\times$, 1.12$\times$, and 1.18$\times$ for the Mixtral-8x7B, Qwen1.5-MoE-A2.7B, and Qwen2-57B-A14B models, respectively. Moreover, HAP attained speedups of up to 1.16$\times$, 1.37$\times$, and 1.11$\times$ on the 4×A100 GPU configuration, demonstrating consistent performance improvements across both architectures when executing Mixtral-series and Qwen-series MoE models. Even in scenarios where TP represents the optimal parallel strategy, HAP achieves comparable execution latency.

\begin{figure*}[htbp]
    \centering
    \includegraphics[width=0.9\linewidth]{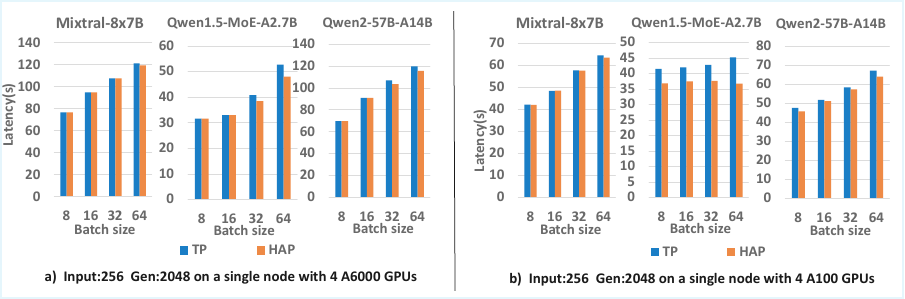}
    \caption{The performance comparison conducted under a scenario featuring a 256-token context and 2048-token generation between HAP and the TP baseline, with evaluations performed on Mixtral-series and Qwen-series MoE models using 4×A6000 GPUs and 4×A100 GPUs, respectively.}
    \label{fig:256-2048}
\end{figure*}
\subsubsection{Short-context input with extended generation.}
As demonstrated in Fig.~\ref{fig:256-2048}, our experimental results reveal distinct acceleration patterns across the three evaluated models. HAP-based inference can achieve at most 1.01$\times$, 1.10$\times$, and 1.036$\times$ speedup on the A6000 GPUs, with corresponding speedups of 1.01$\times$, 1.23$\times$, and 1.05$\times$ observed on A100 GPUs. Notably, HAP frequently fails to surpass the acceleration performance of TP-based inference across multiple configurations. Through systematic analysis, we identify the decoding phase latency as the dominant factor in this particular inference scenario. This observation aligns with our theoretical framework in Section \ref{sec:analysis}, which establishes that the computationally intensive decoding phase inherently favors TP. Consequently, the HAP method consistently prioritizes TP-based configurations for Expert module parallelization during the optimization process, thereby often precluding the attainment of higher acceleration in many scenarios.

\begin{figure*}
    \centering
    \includegraphics[width=0.9\linewidth]{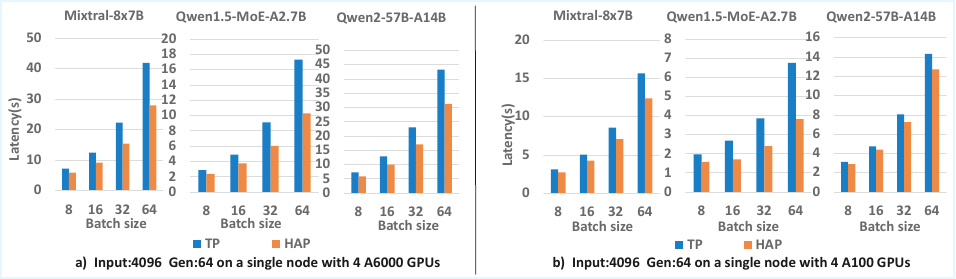}
    \caption{The performance comparison conducted under a scenario featuring a 4096-token context and 64-token generation between HAP and the TP baseline, with evaluations performed on Mixtral-series and Qwen-series MoE models using 4×A6000 GPUs and 4×A100 GPUs, respectively.}
    \label{fig:4096-64}
\end{figure*}

\begin{figure}[htbp]
    \centering
    \includegraphics[width=1\linewidth]{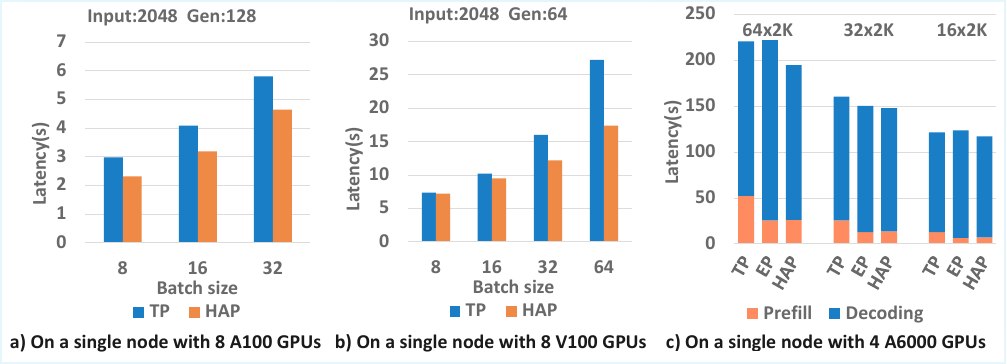}
    \caption{The performance evaluation using Mixtral-8x7B MoE model. (a) The performance comparison under the 2048-token context with 128-token output scenario on a single node with 8 A100 GPUs. (b) The performance comparison under the 2048-token context with 64-token output scenario on a single node with 8 V100 GPUs. (c) The latency comparison between TP-based, EP-based, and HAP-based inference, with evaluations conducted on Mixtral-8x7B model on 4xA6000 GPUS.}
    \label{fig:a100v100}
\end{figure}
\subsubsection{Long-context input with constrained output.}
As illustrated in Fig.~\ref{fig:4096-64}, the HAP-based inference achieves a speedup ranging from 1.21$\times$ to 1.68$\times$ across different MoE models on A6000 GPUs. Similarly, HAP-based inference achieved a maximum speedup of 1.77$\times$ on A100 GPUs, demonstrating the adaptive superiority in specific workload. 
This performance discrepancy stems from the unique computational characteristics of this inference scenario, where the prefill stage consumes a significant amount of time while the decoding stage incurs minimal latency, rendering communication the primary performance bottleneck. In contrast, TP-based inference suffers from excessive communication overhead, highlighting the need for an optimal strategy that minimizes communication to substantially enhance inference efficiency. HAP searched the low-communication parallelization configurations, e.g., Attention module with DP and Expert module with TP or EP, to reduce the communication overhead. 

We also conducted a performance evaluation using the Mixtral-8x7B model under similar inference scenarios featuring a 2048-token input context with constrained output. 
The experiments were carried out on two single-node platforms: one equipped with 8 A100 GPUs, which feature high intra-node bandwidth via NVLINK, and another equipped with 8 NVIDIA V100 GPUs, which rely on limited intra-node bandwidth via PCIe.
As evidenced in the Fig.~\ref{fig:a100v100} (a) and (b), the HAP-based inference yields speedups of of 1.29$\times$ and 1.57$\times$ on A100 and V100 platforms respectively, thereby demonstrating the adaptability of HAP across heterogeneous computing environments.

\begin{figure*}[htbp]
    \centering
    \includegraphics[width=0.9\linewidth]{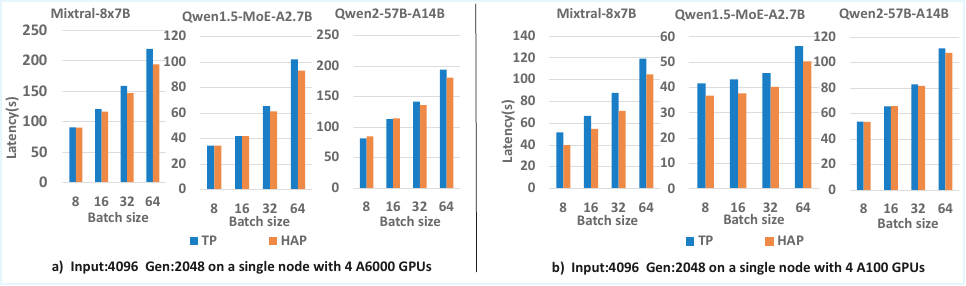}
    \caption{The performance comparison conducted under a scenario featuring a 4096-token context and 2048-token generation between HAP and the TP baseline, with evaluations using 4×A6000 GPUs and 4×A100 GPUs, respectively.}
    \label{fig:4096-2048}
\end{figure*}

\subsubsection{Long-context input with extended output.}
This inference scenario exhibits dual-phase characteristics where decoding latency constitutes the dominant component of end-to-end inference latency, while prefill stage still introduce substantial computational overhead. 
HAP establishes the necessity of phase-specific parallelization strategies: expert parallelism optimizes communication-intensive prefill stage, whereas tensor parallelism accommodate the decoding stage.
As demonstrated in Fig.~\ref{fig:4096-2048}, HAP-based inference achieves up to 1.13$\times$ speedup. 
Moreover, when the prefill stage accounts for a larger proportion of the overall end-to-end latency, the phase-specific parallelization strategies employed by HAP can achieve superior performance.

To substantiate our dynamic parallelism transition methodology, we analyzed the latency in both the prefill and decoding stages across TP-based, EP-based, and HAP-based inference. As depicted in Fig.~\ref{fig:a100v100}(c), EP demonstrates significantly lower prefill latency than TP, albeit with higher decoding latency. In contrast, HAP innovatively integrates dynamic parallelism transition, achieving prefill performance comparable to EP with minimal transition overhead, while simultaneously maintaining decoding efficiency equivalent to TP. This adaptive approach effectively leverages the strengths of both parallelisms.

\vspace{-5pt}
\section{Conclusion}

In this paper, we propose HAP, a novel method that automatically selects hybrid parallel strategies for efficient inference. HAP contains a series of insightful strategies that are theoretically generalizable to different hardware conditions.  
Extensive experiments have demonstrated the excellent efficiency and generalization of HAP on multiple MoE LLMs. 
In future work, we will apply HAP to the multi-node inference, which incorporates a more sophisticated search mechanism.
We will also focus on dynamic, real-time inference serving scenarios to maximize throughput. 


\end{document}